\documentclass[10pt,aps,prx,twocolumn,superscriptaddress,longbibliography,floatfix]{revtex4-2}

\usepackage{amsmath}
\usepackage{amssymb}
\usepackage{graphicx}
\usepackage{dcolumn}
\usepackage{bm}
\usepackage{xcolor}
\usepackage{physics}
\usepackage{microtype}

\usepackage{hyperref}
\hypersetup{
  colorlinks=true,
  linkcolor=blue,
  citecolor=blue,
  urlcolor=blue
}

\begin{document}

\title{Nonclassical Cutoff Fluctuations in Squeezed-Light-Driven High-Harmonic Generation}
% \title{Heisenberg Witness in High-Harmonic Generation via Squeezed Coherent Driving}

\author{Tsendsuren Khurelbaatar}
\email{t.khurelbaatar@griffith.edu.au}
\affiliation{Quantum and Advanced Technologies Research Institute,
Griffith University, Brisbane, Queensland 4111, Australia}
\affiliation{School of Environment and Science,
Griffith University, Brisbane, Queensland 4111, Australia}

\author{R.~T.~Sang}
\affiliation{Quantum and Advanced Technologies Research Institute,
Griffith University, Brisbane, Queensland 4111, Australia}
\affiliation{School of Environment and Science,
Griffith University, Brisbane, Queensland 4111, Australia}
\affiliation{Office of the Pro Vice-Chancellor
(Research, Development and Commercialisation),
University of Southern Queensland, Springfield 4300, Australia}

\author{Igor Litvinyuk}
\affiliation{Quantum and Advanced Technologies Research Institute,
Griffith University, Brisbane, Queensland 4111, Australia}
\affiliation{School of Environment and Science,
Griffith University, Brisbane, Queensland 4111, Australia}

\date{\today}

\begin{abstract}
Strong-field high-harmonic generation (HHG) is conventionally
described semiclassically, with the driving laser treated as a
classical field. This approximation becomes insufficient in nanoscale
interaction geometries, where extreme spatial confinement raises the
vacuum-field amplitude to the $\sim\!10^{-2}$ level relative to the
driving-field amplitude. When the quantum fluctuations of the driving
field are redistributed between conjugate quadratures by squeezing,
they can be directly imprinted onto macroscopic HHG observables. To
model this interaction, we employ a Wigner phase-space approach that
maps the quantum-optical driver onto a stochastic ensemble of
time-dependent Schr\"odinger equation realizations. Although each
realization remains classically simulable, the sub-vacuum quadrature
covariance structure of squeezed states cannot be reproduced by any
field admitting a non-negative Glauber--Sudarshan
$\mathcal{P}$-representation. Within this single-mode Gaussian
framework, we show that amplitude squeezing suppresses the
shot-to-shot variance of the HHG cutoff below the standard quantum
limit (SQL). To leading order in the vacuum-to-driver ratio, the
normalized cutoff variance decays exponentially with the squeezing
parameter, independent of the absolute vacuum-field amplitude and
therefore robust against uncertainties in the effective nanoscale mode
volume. A subleading phase-noise contribution from the anti-squeezed
quadrature produces a variance minimum near $r_{\mathrm{opt}}\approx
1.6$, providing a concrete experimental target. These results
establish the HHG cutoff variance ratio as a nonlinear,
self-calibrating Heisenberg witness in which sub-SQL scaling directly
reflects the redistribution of quantum uncertainty in the driving
field.
\end{abstract}

\maketitle

%===================================================================
\section{Introduction}
\label{sec:intro}
%===================================================================
Attosecond science fundamentally rests on high-harmonic generation (HHG), a strongly nonperturbative process accurately captured by the semiclassical three-step model~\cite{corkum1993,lewenstein1994}. In the conventional treatment of this process, the driving laser is assumed to be a deterministic classical field, rendering photon statistics operationally irrelevant. This approximation is well-justified in free-space focusing geometries, where the large quantization volume heavily suppresses the vacuum-field amplitude, rendering quantum fluctuations negligible relative to the amplitude of the driving field. Built upon this deterministic foundation, HHG has successfully underpinned isolated attosecond pulse generation~\cite{hentschel2001,paul2001}, nonsequential double ionization~\cite{walker1994precision,fittinghoff1992observation}, and a broad spectrum of nonperturbative laser--matter phenomena.

Recent developments in quantum light sources, however, are beginning to challenge this established paradigm. Bright squeezed vacuum~\cite{iskhakov2012,chekhova2015,perez2015} and related nonclassical fields are now being used to drive strong-field processes, establishing strong-field quantum optics as a rapidly evolving experimental frontier~\cite{rasputnyi2024,lemieux2025,heimerl2025,Sennaryetal2025}. Parallel theoretical work on quantum-optical driving~\cite{Gorlach2023,Bhattacharya2023,Tzur2023,EvenTzur2023PhotonStatistics} has predicted phenomena that extend beyond the semiclassical framework, including the generation of Schr\"odinger cat-states during laser--matter interactions~\cite{lewenstein_generation_2021}. Despite this progress, a central question remains unresolved: under what specific conditions do quantum field statistics manifest in macroscopic strong-field observables, and how can their unique signatures be distinguished from classical stochastic noise?

The standard reference for making this distinction is the standard quantum limit (SQL), which represents the coherent-state shot-noise floor. The decisive criterion for nonclassicality is the non-negativity of the Glauber--Sudarshan $\mathcal{P}$-representation~\cite{glauber1963,sudarshan1963}. Any field with $\mathcal{P}\!\ge\!0$ is strictly a statistical
mixture of coherent states, meaning every marginal quadrature
variance is bounded below by the vacuum value
($\sigma_{X_\theta}^{2}\!\ge\!1/2$ for any rotated quadrature
$X_\theta = X\cos\theta + P\sin\theta$). Consequently, observing sub-SQL fluctuations in any single quadrature serves as a rigorous certification of nonclassicality. To date, however, no strong-field observable has been identified that provides an experimentally accessible, sub-SQL operational witness.

Here, we identify such an observable. We show that the HHG cutoff variance,
when normalized to its coherent-state SQL value, acts as a self-calibrating
Heisenberg witness. By “Heisenberg witness,” we mean an observable whose sub-SQL scaling certifies the redistribution of uncertainty between conjugate quadratures imposed by the Heisenberg constraint.
Its leading-order scaling is dictated entirely by the squeezing parameter
and remains independent of the effective quantization volume
$V_{\mathrm{eff}}$, which is typically the largest source of theoretical
uncertainty in nanoscale geometries. The effect is most pronounced in nanoscale interaction volumes---such as nanoplasmonic hotspots, metallic nanotips, semiconductor nanostructures, and dielectric metasurfaces~\cite{kim_high-harmonic_2008,ciappina2017,kruger2011,sivis2017,vampa2017}. In these systems, spatial confinement reduces the quantization volume of the driving field to the scale of the electron excursion, thereby elevating the vacuum-field amplitude and allowing quantum fluctuations to reach the $\sim\!10^{-2}$ level relative to the driving field amplitude (as detailed in Sec.~\ref{sec:wigner}).

An enhanced vacuum-field amplitude alone is insufficient: the resulting isotropic vacuum fluctuations merely broaden the field distribution without producing quadrature-selective effects. The critical ingredient is instead the constrained redistribution of these vacuum fluctuations between conjugate quadratures, a direct consequence of the Heisenberg uncertainty relation, whereby suppression in one quadrature necessitates enhancement in the other. 

\begin{figure*}[htbp]
  \centering
  \includegraphics[width=0.8\linewidth]{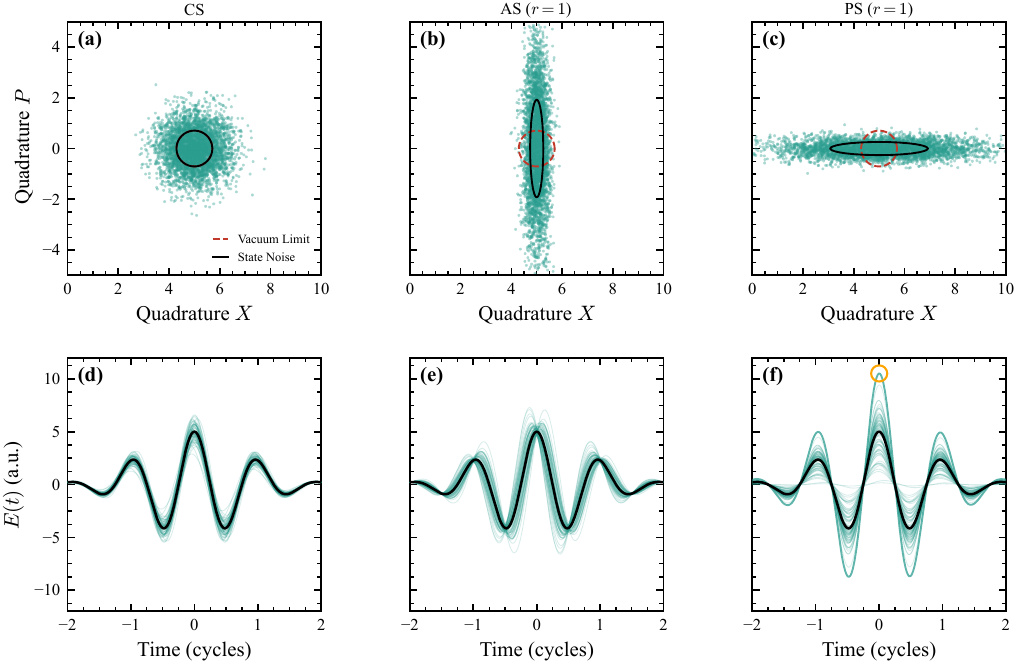}
    \caption{\textbf{Wigner phase-space representation and stochastic
    field realizations.} Top row: Wigner distributions for (a)~a
    coherent state (CS), (b)~an amplitude-squeezed state (AS, $r=1$),
    and (c)~a phase-squeezed state (PS, $r=1$). Teal points are
    quadrature pairs $(X_i, P_i)$ sampled from the corresponding Wigner
    distribution; solid black ellipses are the $1\sigma$ contours, and
    dashed crimson circles indicate the SQL. Bottom row: time-domain
    field realizations $E(t)$ for (d)~coherent, (e)~amplitude-squeezed,
    and (f)~phase-squeezed pulses. Thin teal lines show individual
    stochastic realizations; the solid black line is the mean field. In
    (f), a rare high-amplitude realization (highlighted in orange)
    illustrates the statistical fluctuations responsible for the HHG cutoff extension.}
    \label{fig:field_statistics}
\end{figure*}

To systematically model this, we treat the problem using a Wigner phase-space formulation, which maps the quantum-optical driving exactly onto a stochastic ensemble of time-dependent Schr\"odinger equation (TDSE) realizations~\cite{cahill1969}. Because squeezed coherent states possess positive Gaussian Wigner distributions, this mapping is exact. This yields two major consequences. First, the realizations themselves are classically simulable, meaning the dynamics admit a sample-by-sample classical stochastic representation. Second, squeezed states possess a sub-vacuum quadrature covariance structure that cannot be reproduced by any single-mode classical stochastic field with a non-negative $\mathcal P$-representation. Such states are admissible because the Heisenberg bound $\sigma_X\sigma_P\ge 1/2$ permits minimum-uncertainty states with $\sigma_X^2<1/2$, provided the conjugate quadrature variance is correspondingly enhanced. The distinguishing quantum feature, therefore, is rooted in the covariance structure of the admissible field statistics rather than in the realization-level dynamics themselves.

Based on this framework, our analysis yields three primary results. First, within the single-mode Gaussian framework, amplitude squeezing suppresses the HHG cutoff variance $\Delta H^{2}$ below the SQL. We find that the leading-order behavior of the ratio $\Delta H^{2}/\Delta H^{2}_{\mathrm{SQL}}=e^{-2r}$ is independent of the absolute vacuum-field amplitude. Second, we observe that the variance depends continuously on the squeezing angle $\theta$, mirroring the orientation of the Wigner ellipse in phase space. Third, we identify a subleading phase-noise channel driven by the conjugate quadrature variance ($\sigma_P^{2}\!\propto\!e^{+2r}$). This channel produces a distinct variance minimum at $r_{\mathrm{opt}}\!\approx\!1.6$ within our saddle-point evaluation, illustrating a measurable competition between conjugate-quadrature noise contributions to the same macroscopic observable. Ultimately, the robust normalization of the HHG cutoff variance ratio provides a Heisenberg witness that is protected against uncertainties in $V_{\mathrm{eff}}$.

%===================================================================
\section{Theoretical Framework}
\label{sec:theory}
%===================================================================

\subsection{Squeezed coherent states}
\label{sec:squeezing}

To capture the interplay between quantum statistics and strong-field dynamics, we first define the properties of the displaced squeezed states driving the interaction, denoted $\ket{\alpha,\xi}=\hat{D}(\alpha)\hat{S}(\xi)\ket{0}$. Following standard conventions~\cite{walls1983,loudon1987}, the displacement is applied after the squeezing operator. The complex displacement $\alpha=|\alpha|e^{i\phi}$ specifies the mean coherent amplitude and optical phase, while the squeezing parameter $\xi=re^{i\theta}$ determines the magnitude $r\!\ge\!0$ and angle $\theta$ of the quadrature redistribution. In our notation, $r\!>\!0$ with $\theta=0$ represents amplitude squeezing (AS), which suppresses fluctuations of the in-phase quadrature $X$ below the vacuum level. Conversely, $\theta=\pi/2$ represents phase squeezing (PS), suppressing fluctuations of the orthogonal quadrature $P$.

By definition, these states saturate the Heisenberg uncertainty relation,
\begin{equation}
  \sigma_X \sigma_P = \tfrac{1}{2},
  \label{eq:heisenberg}
\end{equation}
where $\sigma_X$ and $\sigma_P$ represent the standard deviations of the dimensionless quadratures. Atomic units ($ \hbar = m_e = e = a_0 = 1 $) are used throughout unless otherwise stated. Squeezing effectively redistributes a fixed quantum uncertainty budget between the conjugate quadratures while preserving this minimum-uncertainty product. When viewed in phase space, this transformation stretches the isotropic circular profile of a coherent state into an ellipse of equal area, as illustrated in Figs.~\ref{fig:field_statistics}(a)--(c).

For states aligned with the principal quadratures, these variances scale exponentially with $r$. Under the AS convention, we have
\begin{equation}
  \sigma_X^{2}(r)=\tfrac{1}{2}e^{-2r},
  \qquad
  \sigma_P^{2}(r)=\tfrac{1}{2}e^{+2r},
  \label{eq:variances}
\end{equation}
with the assignments interchanging for PS. The anti-squeezed quadrature becomes important when
channeled through the highly nonlinear Ammosov--Delone--Krainov (ADK)
equation~\cite{adk1986}: even modest quadrature
redistribution is amplified, resulting in measurable
modifications to the strong-field response.

%===================================================================

\subsection{Wigner phase-space mapping and vacuum-field amplitude}
\label{sec:wigner}

To connect these quantum states to the strong-field electron response, we utilize the Wigner phase-space formulation. Because squeezed coherent states possess positive Gaussian Wigner distributions, quantum expectation values map exactly onto stochastic classical averages~\cite{cahill1969,schleich2001quantum,gardiner2004quantum}. This effectively reduces the driven quantum-optical strong-field problem to an ensemble of independent stochastic field realizations, each appropriately weighted by the Wigner distribution. Crucially, this mapping is exact rather than a semiclassical approximation, meaning the realization-level dynamics are classically simulable. The divergence from classicality occurs only at the level of admissible covariance: any single-mode classical stochastic field satisfying $\mathcal{P}\!\ge\!0$ must have a marginal quadrature variance bounded below by $1/2$ in any rotated direction, a bound that squeezed states violate by design.

When squeezing is aligned with the principal quadratures ($\theta=0$ or $\pi/2$), the Wigner distribution takes the form of a diagonal Gaussian,
\begin{equation}
  W(X,P)=\frac{1}{2\pi\sigma_X\sigma_P}
  \exp\!\left[
    -\frac{(X-X_c)^{2}}{2\sigma_X^{2}}
    -\frac{P^{2}}{2\sigma_P^{2}}
  \right],
  \label{eq:wigner}
\end{equation}
where $X_c=|\alpha|$ and $P_c=0$. To accommodate arbitrary angles $\theta$, the distribution naturally generalizes to a rotated Gaussian with an off-diagonal covariance structure:
\begin{align}
  \sigma_X^{2}(\theta)
  &= \tfrac{1}{2}\!\left(\cos^{2}\theta\,e^{-2r}+\sin^{2}\theta\,e^{+2r}\right),\\
  \sigma_P^{2}(\theta)
  &= \tfrac{1}{2}\!\left(\sin^{2}\theta\,e^{-2r}+\cos^{2}\theta\,e^{+2r}\right),\\
  \sigma_{XP}(\theta)
  &= \tfrac{1}{2}\sin\theta\cos\theta\!\left(e^{+2r}-e^{-2r}\right).
\end{align}
This rotated covariance formulation underlies the $\theta$-dependent simulations discussed in Sec.~\ref{sec:quadrature_signature}. As expected, the cross-covariance vanishes entirely at $\theta=0$ and $\pi/2$.

By performing Monte Carlo sampling of Eq.~\eqref{eq:wigner}, we generate an ensemble of quadrature pairs $(X_i,P_i)$ that inherently obey the Heisenberg constraint. Each sampled pair defines a deterministic time-domain field realization,
\begin{equation}
  E_i(t)
  =
  \mathcal{E}_{\mathrm{vac}}
  \exp\!\left[
    -2\ln 2
    \left(\frac{\omega t}{2\pi N}\right)^{\!2}
  \right]
  \!\left(X_i\cos\omega t+P_i\sin\omega t\right),
  \label{eq:field}
\end{equation}
where $\omega$ is the carrier frequency, $N$ is the pulse duration in optical cycles, and the vacuum-field amplitude is defined as
\begin{equation}
\mathcal{E}_{\mathrm{vac}} = \sqrt{\frac{2\pi\omega}{V_{\mathrm{eff}}}}.
\label{eq:Evac}
\end{equation}
Representative time-domain realizations generated from Eq.~\eqref{eq:field} are shown in Figs.~\ref{fig:field_statistics}(d)--(f). For the coherent state [Fig.~\ref{fig:field_statistics}(d)], individual shots fluctuate about the mean field at the SQL level. Amplitude squeezing [Fig.~\ref{fig:field_statistics}(e)] suppresses amplitude fluctuations near the field extrema, while the conjugate phase quadrature is correspondingly broadened. Conversely, phase squeezing [Fig.~\ref{fig:field_statistics}(f)] stabilizes the phase but substantially broadens the amplitude distribution. A rare high-amplitude realization (highlighted in orange) illustrates the statistical outliers responsible for the above-cutoff HHG extension discussed in Sec.~\ref{sec:cutoff_extension}.

The physical geometry of the nanostructure is encoded entirely within
$V_{\mathrm{eff}}$ through Eq.~\eqref{eq:Evac}. As a representative
order-of-magnitude estimate (evaluated in SI units), setting
$V_{\mathrm{eff}}\!\sim\!(\lambda/300)^{3}$ at a wavelength of
$\lambda=1500$~nm yields $\mathcal{E}_{\mathrm{vac}}\!\sim\!
2.4\times10^{8}~\mathrm{V/m}$. In contrast, a peak intensity of
$I_0=10^{14}~\mathrm{W/cm^{2}}$ produces a macroscopic peak field of
$E_0\!\sim\!2.7\times10^{10}~\mathrm{V/m}$. The resulting ratio
$\mathcal{E}_{\mathrm{vac}}/E_0\!\sim\!10^{-2}$ is consistent with
tightly confined nanoscale regimes currently accessible in
state-of-the-art experimental platforms~\cite{kim_high-harmonic_2008,
ciappina2017,sivis2017,vampa2017}. Three independent atomic-unit-based
estimates confirm that
this ratio is physically well-motivated for near-field nanoplasmonic
geometries as detailed in the SM~\cite{SM}.

%===================================================================

\subsection{Stochastic TDSE and the cutoff observable}
\label{sec:stochastic_tdse}

With the stochastic field realizations defined, we evaluate the strong-field electron dynamics. For each realization $E_i(t)$ drawn from Eq.~\eqref{eq:field}, we numerically solve the one-dimensional length-gauge TDSE for an argon model atom ($I_p\!\approx\!15.8~\mathrm{eV}$, modeled via a soft-core potential~\cite{su1991model,javanainen1988numerical}). The target is driven by a linearly polarized two-cycle pulse at $\lambda=1500~\mathrm{nm}$ with a peak intensity $I_0=10^{14}~\mathrm{W/cm^{2}}$, placing the dynamics in the tunneling regime with the Keldysh parameter $\gamma\!\lesssim\!0.5$~\cite{keldysh1965}. We utilize a 1D atomic target to provide a controlled platform for testing our statistical predictions; the nanoscale geometry enters the problem exclusively through the parameter $\mathcal{E}_{\mathrm{vac}}$. Comprehensive numerical details and convergence tests are available in the SM~\cite{SM}.

The HHG spectrum for each individual realization is calculated from the Fourier transform of the Ehrenfest dipole acceleration,
\begin{equation}
  a(t)
  =
  \left\langle
    -\frac{\partial V}{\partial x}
    -E(t)
  \right\rangle,
\end{equation}
following the standard approach of Refs.~\cite{burnett1992,baggesen2011}. The resulting spectral intensity is given by
\begin{equation}
  S(\omega)
  =
  \left|
    \int_{-\infty}^{\infty}
    a(t)\,W(t)\,e^{-i\omega t}\,\mathrm{d}t
  \right|^{2},
  \label{eq:hhg_spectrum}
\end{equation}
where $W(t)$ is an applied Blackman window. From each calculated $S_i(\omega)$, we extract a per-shot cutoff position $H_i$. This is defined as the spectral position where the smoothed $\log S_i(\omega)$ curve drops a predetermined number of decades below its characteristic plateau value (the exact extraction protocol is detailed in the SM~\cite{SM}). The statistical mean $\langle H\rangle$ and variance $\Delta H^{2} = \langle H^{2}\rangle - \langle H\rangle^{2}$ are calculated as ensemble averages over $N_{\mathrm{shot}}=1000$ distinct realizations. 

Crucially, $\Delta H^{2}$ is a directly extractable quantity from standard shot-to-shot HHG measurements, requiring neither homodyne detection nor photon-counting capabilities. Because the variance signatures derived below originate fundamentally from the statistics of the driving field, they are expected to remain physically robust beyond the present 1D implementation. We provide validation of this robustness against deterministic TDSE simulations in the SM~\cite{SM}.

\begin{figure}[htbp]
  \centering
  \includegraphics[width=\linewidth]{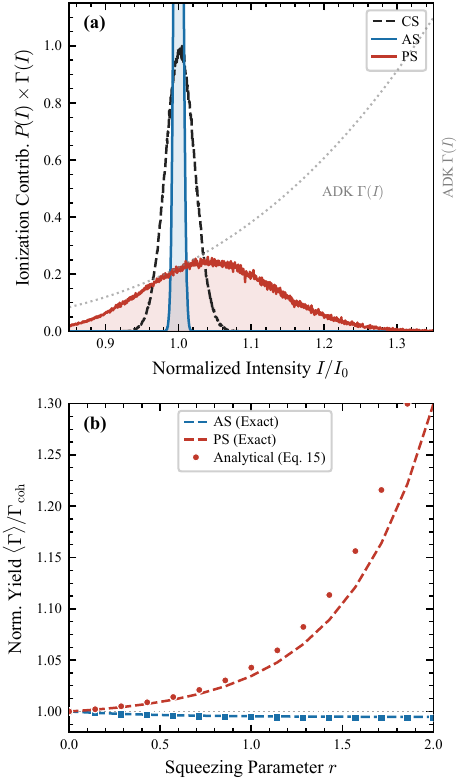}
    \caption{\textbf{Statistical filtering of tunnel ionization.}
    (a)~Ionization contribution $P(I)\Gamma(I)$ versus normalized
    intensity $I/I_0$ for CS (black dashed), AS (blue), and PS (red
    shaded). The bare ADK rate $\Gamma(I)$ (gray dotted, right axis)
    illustrates exponential weighting toward rare high-intensity
    realizations. (b)~Normalized ionization yield
    $\langle\Gamma\rangle/\Gamma_{\mathrm{coh}}$ versus $r$, comparing
    numerical evaluation of Eq.~\eqref{eq:ensemble} for AS (blue dashed)
    and PS (red dashed) with the analytical approximation
    Eq.~\eqref{eq:net_enhancement} (filled symbols).}
    \label{fig:ionization}
\end{figure}
%===================================================================
\section{Results}
\label{sec:results}
%===================================================================

\subsection{Ionization and statistical filtering}
\label{sec:adk_expansion}

In the tunneling regime ($\gamma\!<\!1$)~\cite{keldysh1965}, the instantaneous ionization rate is highly nonlinear and is well-described by the ADK formula~\cite{adk1986}:
\begin{equation}
  \Gamma(E)
  \sim
  A(E)
  \exp\!\left(-\frac{B}{|E|}\right),
  \label{eq:adk}
\end{equation}
where $B=\tfrac{2}{3}(2I_p)^{3/2}$ and $A(E)$ is a weakly field-dependent prefactor. The macroscopic, observable yield is the convolution of this rate with the field statistics:
\begin{equation}
  \langle\Gamma\rangle
  =
  \int_{-\infty}^{\infty}
  dE\;Q(E)\Gamma(E),
  \label{eq:ensemble}
\end{equation}
where $Q(E)$ represents the marginal field distribution at the pulse maximum.

The steep exponential structure of Eq.~\eqref{eq:adk} effectively acts as a statistical filter. Rare, high-amplitude field realizations [e.g., see Fig.~\ref{fig:field_statistics}(f)] are weighted exponentially more strongly than fluctuations occurring near the mean field, as visualized in Fig.~\ref{fig:ionization}(a). By expanding $\ln\Gamma(E)$ as a Taylor series about the mean field $E_0$,
\begin{equation}
  \ln\Gamma(E)
  =
  \ln\Gamma(E_0)
  +
  \frac{B}{E_0^{2}}\delta E
  -
  \frac{B}{E_0^{3}}\delta E^{2}
  +
  \mathcal{O}(\delta E^{3}),
  \label{eq:expansion}
\end{equation}
and applying a cumulant expansion tailored to the Gaussian field distribution, we can isolate the leading-order yield as
\begin{equation}
  \ln\frac{\langle\Gamma\rangle}{\Gamma(E_0)}
  \simeq
  \mathcal{E}_{\mathrm{vac}}^{2}\sigma_X^{2}(r)\!\left(\frac{B^{2}}{2E_0^{4}}-\frac{B}{E_0^{3}}\right).
  \label{eq:cumulant}
\end{equation}
The bracketed coefficient captures two competing effects: a positive enhancement arising from the squared linear term in Eq.~\eqref{eq:expansion}, and a partial cancellation from the negative quadratic correction. For the physical parameters considered here, the first term dominates, resulting in a net yield enhancement whenever amplitude statistics are broadened. If we define the dimensional vacuum field variance $\sigma_{\mathrm{vac}}^{2} = \mathcal{E}_{\mathrm{vac}}^{2}/2$ and a dimensionless parameter $\eta = \sigma_{\mathrm{vac}}^{2}\,(B^{2}/2E_0^{4}-B/E_0^{3})$, the effective ensemble yield, normalized to the coherent-state yield ($\Gamma_{\mathrm{coh}}$), simplifies to
\begin{equation}
  \frac{\langle\Gamma\rangle}{\Gamma_{\mathrm{coh}}}
  \simeq
  \exp\!\left[\eta\!\left(e^{-2r}-1\right)\right].
  \label{eq:net_enhancement}
\end{equation}

Figure~\ref{fig:ionization}(b) demonstrates that this statistical enhancement is sensitive to the specific quadrature variance. PS states exhibit a monotonic yield enhancement due to their broadened amplitude statistics. Conversely, AS states produce little modification; their suppressed amplitude fluctuations are transferred into the phase quadrature, which does not couple at leading order to the tunneling exponent. We emphasize that this enhancement is a statistical consequence, not an exclusively quantum one: any classical field possessing an identical marginal amplitude distribution would generate the same yield response. The role of the squeezed state here is to provide a rigorously Heisenberg-bounded, experimentally tunable mechanism for shaping single-shot field statistics. Quantum signatures emerge when examining fluctuations of the HHG cutoff.

%===================================================================

\subsection{HHG cutoff extension}
\label{sec:cutoff_extension}

\begin{figure}[htbp]
  \centering
  \includegraphics[width=\columnwidth]{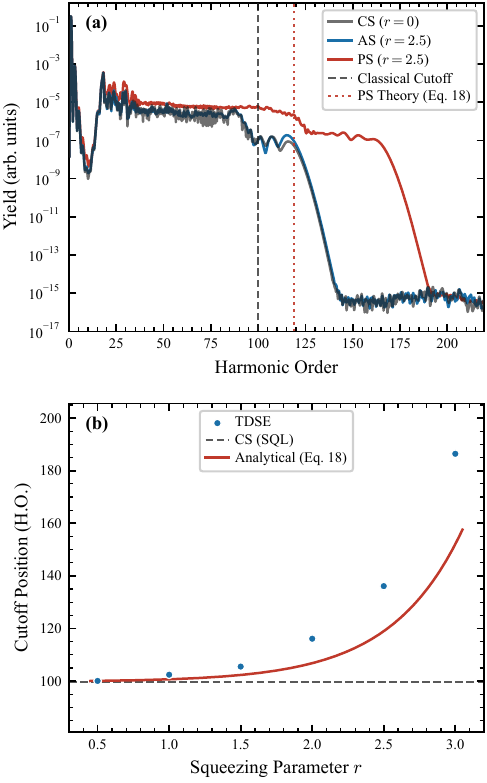}
    \caption{\textbf{HHG cutoff extension under squeezed-field driving.}
    (a)~Ensemble-averaged HHG spectra for CS (black), AS (blue), and PS
    (red) drivers at $r=2.5$. The classical cutoff
    $I_p+3.17U_p(E_0)$ (gray dashed) and the analytical stochastic
    prediction Eq.~\eqref{eq:s_cutoff} (red dotted, PS Theory) are
    indicated. (b)~Extracted cutoff position versus $r$ from TDSE
    simulations (filled circles) and from the analytical prediction
    Eq.~\eqref{eq:s_cutoff} (red solid) for the phase-squeezed (PS) case, with the coherent-state SQL
    reference (gray dashed).}
    \label{fig:hhg_spectrum}
\end{figure}

The statistical filtering of the ionization step fundamentally shapes the macroscopic HHG spectrum. Figure~\ref{fig:hhg_spectrum}(a) displays the ensemble-averaged HHG spectra computed at a fixed mean field $E_0$. When compared to the coherent-state reference, the PS driver generates a significant above-cutoff extension spanning several tens of harmonic orders (H.O.), whereas the AS spectrum remains close to the classical limit. This high-energy extension is a direct macroscopic consequence of the statistical filtering detailed in Sec.~\ref{sec:adk_expansion}: a small subset of high-amplitude stochastic realizations contributes disproportionately to the ensemble, generating larger ponderomotive energies and pushing the rate-weighted cutoff beyond the mean-field prediction. Each individual realization strictly obeys the standard classical cutoff law,
\begin{equation}
  E_{\mathrm{cutoff}}=I_p+3.17U_p(E),
  \label{eq:classical_cutoff}
\end{equation}
evaluated at its own instantaneous field amplitude. The observed extension emerges upon statistical averaging over the ensemble.

Mathematically, the ensemble cutoff behaves as the rate-weighted average
\begin{equation}
  \langle E_{\mathrm{cutoff}}\rangle
  =
  I_p
  +
  3.17
  \frac{\langle\Gamma(E)U_p(E)\rangle}{\langle\Gamma(E)\rangle}.
  \label{eq:rate_weighted}
\end{equation}
By expanding $U_p(E)$ about $E_0$ and integrating this with the previously derived cumulant expansion of the ADK rate, we arrive at an analytical expression for the cutoff shift:
\begin{equation}
  E_{\mathrm{cutoff}}(r) \simeq I_p + 3.17 U_p(E_0) \left[ 1 + \epsilon^2 \left(1+\frac{2B}{E_0}\right) \right],
  \label{eq:s_cutoff}
\end{equation}
which is valid in the $\epsilon \ll 1$ limit. Here, the small parameter $\epsilon$ is defined as
\begin{equation}
  \epsilon = \frac{\mathcal{E}_{\mathrm{vac}}\,e^{r}}{\sqrt{2}\,E_0},
  \label{eq:epsilon}
\end{equation}
and dictates the magnitude of the largest quadrature fluctuation relative to the mean driving field. All analytical expressions in this work are asymptotic in this parameter. As shown in Fig.~\ref{fig:hhg_spectrum}(b), the analytical model captures both the magnitude and the monotonic scaling of the TDSE-extracted cutoffs at moderate squeezing levels (e.g., at $r=1.5$, $\epsilon \sim 3\times10^{-2}$). Predictable deviations emerge at larger $r$ (such as $r=3.0$, where $\epsilon \sim 1.4\times10^{-1}$), as higher-order cumulants become non-negligible.

\begin{figure*}[htbp]
    \centering
    \includegraphics[width=0.8\textwidth]{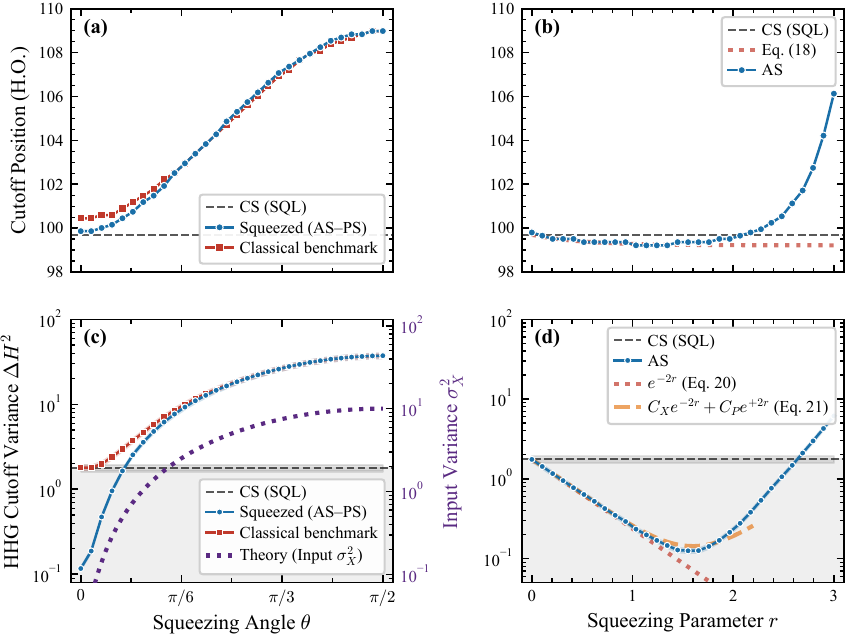}
    \caption{\textbf{Quadrature-resolved HHG cutoff statistics.}
    (a)~Mean cutoff $\langle H\rangle$ versus squeezing angle $\theta$ at fixed $r=1.5$ for the squeezed state (blue circles) and the classical $\mathcal{P}\!\ge\!0$ benchmark (dark red squares). The CS SQL reference is shown as a gray dashed line.
    (b)~Mean cutoff versus $r$ under amplitude squeezing ($\theta=0$) for AS (blue circles), the CS SQL reference (gray dashed), and the analytical prediction from Eq.~\eqref{eq:s_cutoff} (red dotted curve). The exact TDSE diverges upward from the analytical model at extreme squeezing due to higher-order phase noise contributions.
    (c)~HHG cutoff variance $\Delta H^{2}$ versus $\theta$. The squeezed state (blue circles) descends well below the SQL at $\theta = 0$ and remains sub-SQL over an extended angular range before crossing the coherent-state reference; the classical benchmark (dark red squares) remains bounded by the SQL across all $\theta$. The right axis and purple dotted curve show the corresponding input quadrature variance $\sigma_X^2(\theta)$, confirming that $\Delta H^2$ tracks $\sigma_X^2(\theta)$ across the full angular sweep.
    (d)~$\Delta H^{2}$ versus $r$ under amplitude squeezing. The leading-order $e^{-2r}$ scaling [Eq.~\eqref{eq:ratio}] is shown as a red dotted curve. The two-channel analytical model [Eq.~\eqref{eq:two_channel}, orange dashed curve], plotted up to $r=2.2$ where the small-parameter expansion remains valid, accurately captures the variance minimum at $r_{\mathrm{opt}}\!\approx\!1.6$.}
    \label{fig:quantum_advantage}
\end{figure*}

While the mean cutoff extension demonstrates sensitivity to
the underlying field statistics, it does not intrinsically
establish nonclassicality: an engineered classical stochastic
field with matched amplitude statistics would reproduce an
identical mean cutoff shift. To find a decisive distinction,
we must look to the fluctuations of the HHG cutoff.
%===================================================================

\subsection{Quadrature-resolved cutoff statistics}
\label{sec:quadrature_signature}

To establish a classicality criterion, we recall that a single-mode field is considered classical in the quantum-optical sense if it admits a positive Glauber--Sudarshan representation ($\mathcal{P}\!\ge\!0$)~\cite{glauber1963,sudarshan1963}. Because such fields are statistical mixtures of coherent states, their quadrature covariance decomposes as $\Sigma = \Sigma_{\mathrm{vac}} + \Sigma_{\mathrm{cl}}$, where $\Sigma_{\mathrm{vac}} = 1/2$ and $\Sigma_{\mathrm{cl}} \ge 0$. A corollary of this is that every marginal variance---including measurements along any arbitrarily rotated quadrature $X_\theta = X\cos\theta + P\sin\theta$---remains bounded strictly below by $1/2$. This holds true even in the presence of classical cross-correlations ($\sigma_{XP}^{\mathrm{cl}}\!\ne\!0$). The threshold certification for nonclassicality (i.e., achieving sub-SQL marginal variance) can therefore be applied rigorously along any selected quadrature direction.

With this in mind, we construct a classical benchmark to serve as the most permissive single-mode model consistent with $\mathcal{P}\!\ge\!0$. In this benchmark, we allow full freedom to identically match the squeezed-state amplitude variance $\sigma_X^{2}(r,\theta)$, while fixing the conjugate variance at $\sigma_P^{2}=1/2$ to saturate the SQL bound from below. Because introducing classical correlations $\sigma_{XP}^{\mathrm{cl}}\!\ne\!0$ cannot push a marginal variance below the vacuum limit, choosing $\sigma_{XP}^{\mathrm{cl}}=0$ does not artificially weaken the comparison. Any alternative classical field with $\sigma_P^{2}\!<\!1/2$ would inherently certify its own nonclassicality, and any with $\sigma_P^{2}\!>\!1/2$ would represent a more conservative bound. Our benchmark therefore represents the tightest single-mode classical comparison available, isolating the impact of the covariance structure imposed by the strict saturation $\sigma_X\sigma_P=1/2$.

When examining the mean cutoff behavior, we find it is classically reproducible. Figure~\ref{fig:quantum_advantage}(a) plots $\langle H\rangle$ versus $\theta$ at a fixed $r=1.5$. The quantum and classical results are nearly indistinguishable, consistent with our finding from Eq.~\eqref{eq:s_cutoff} that the mean shift relies exclusively on the marginal amplitude variance.

The fluctuations, however, distinguish between the two regimes [Fig.~\ref{fig:quantum_advantage}(c)]. The classical benchmark remains SQL-bounded across all values of $\theta$, whereas the squeezed state descends below the SQL, reaching a minimum at $\theta=0$. By linearizing the classical cutoff relationship [Eq.~\eqref{eq:classical_cutoff}] about $E_0$, we obtain $\delta H \simeq (\partial H/\partial E)\,\delta E$. The derivative term $(\partial H/\partial E)=2(3.17 U_p(E_0)/E_0)$ is identical for both quantum and classical drivers evaluated at the same $E_0$. The resulting variance is $\Delta H^{2} = (\partial H/\partial E)^{2}\sigma_{\delta E}^{2}$, with $\sigma_{\delta E}^{2} = \mathcal{E}_{\mathrm{vac}}^{2}\sigma_X^{2}(r,\theta=0)$ in the case of amplitude squeezing. By normalizing this expression against the coherent-state variance at the same mean field, both the linear-response prefactor and the absolute vacuum-field amplitude cancel at leading order:
\begin{equation}
  \frac{\Delta H^{2}(r)}{\Delta H_{\mathrm{SQL}}^{2}}
  =
  \frac{\sigma_X^{2}(r,\theta=0)}{\sigma_{X,\mathrm{SQL}}^{2}}
  =
  e^{-2r}.
  \label{eq:ratio}
\end{equation}
This cancellation is exact at $\mathcal{O}(\epsilon^{0})$. 
Conversely, phase squeezing ($\theta=\pi/2$) yields
$\Delta H^{2}/\Delta H_{\mathrm{SQL}}^{2}=e^{+2r}>1$,
consistent with the super-SQL behavior visible in
Fig.~\ref{fig:quantum_advantage}(c).
While explicit dependence on $\mathcal{E}_{\mathrm{vac}}$ does re-emerge at higher orders (particularly when incorporating the phase-noise channel discussed below), to leading order, Eq.~\eqref{eq:ratio} is independent of $V_{\mathrm{eff}}$. This constitutes the central result of our work: we have isolated a normalized Heisenberg witness whose scaling behavior is dictated entirely by the squeezing parameter and is protected against the dominant theoretical uncertainties associated with nanoscale geometries. Furthermore, across the full angular sweep, $\Delta H^{2}$ tracks $\sigma_X^{2}(\theta)$, yielding a quadrature-resolved mapping between the abstract Wigner ellipse and the physical cutoff observable.

To ensure a complete physical picture, we must address the subleading phase-noise channel. Figures~\ref{fig:quantum_advantage}(b) and (d) chart $\langle H\rangle$ and $\Delta H^{2}$ against $r$ specifically under amplitude squeezing. While $\Delta H^{2}$ closely follows Eq.~\eqref{eq:ratio} for $r\!\lesssim\!1$, it develops a minimum near $r_{\mathrm{opt}}\!\approx\!1.6$. To model this turnaround, we must include the conjugate quadrature in our linearization. At $\theta=0$, the cross-covariance vanishes, rendering amplitude and phase fluctuations statistically independent. By writing $\delta H \simeq (\partial H/\partial E)\,\delta E + (\partial H/\partial t_{\mathrm{ion}})\,\delta t_{\mathrm{ion}}$, the variance separates into two components:
\begin{equation}
  \Delta H^{2}(r)
  =
  C_X\,e^{-2r}
  +
  C_P\,e^{+2r},
  \label{eq:two_channel}
\end{equation}
with the respective coefficients defined as
\begin{equation}
  C_X
  =
  \tfrac{1}{2}\mathcal{E}_{\mathrm{vac}}^{2}
  \!\left(\frac{\partial H}{\partial E}\right)^{\!2},
  \quad
  C_P
  =
  \tfrac{1}{2}
  \!\left(\frac{\mathcal{E}_{\mathrm{vac}}}{\omega E_0}\right)^{\!2}
  \!\left(\frac{\partial H}{\partial t_{\mathrm{ion}}}\right)^{\!2}.
  \label{eq:CXP_coef}
\end{equation}
Here, the factor $\mathcal{E}_{\mathrm{vac}}/(\omega E_0)$ effectively translates a fluctuation in the $P$-quadrature into a corresponding ionization-time jitter via the relation $\delta t_{\mathrm{ion}}\simeq -\mathcal{E}_{\mathrm{vac}}\delta P/(\omega E_0)$, evaluated at the saddle-point ionization time. The subscripts $X$ and $P$ identify which quadrature variance each coefficient couples to. Both partial derivatives are evaluated at the deterministic saddle 
point of the three-step model; their explicit forms and the 
saddle-point evaluation of $r_{\mathrm{opt}}$ are derived in 
the SM~\cite{SM}.

Minimizing this two-channel variance yields an optimal squeezing parameter $r_{\mathrm{opt}}=\tfrac{1}{4}\ln(C_X/C_P)$. Because the common $\mathcal{E}_{\mathrm{vac}}^{2}$ factors cancel out, the ratio
\begin{equation}
  \frac{C_X}{C_P}
  =
  \omega^{2} E_0^{2}
  \!\left(\frac{\partial H/\partial E}{\partial H/\partial t_{\mathrm{ion}}}\right)^{\!2}
  \label{eq:CXP_ratio}
\end{equation}
is mode-volume-independent. Thus, the physical location of the optimum is dictated by the strong-field response dynamics rather than the specific parameters of the nanostructure geometry. While the precise location of this minimum is model-dependent (relying heavily on the saddle-point evaluation of $\partial H/\partial t_{\mathrm{ion}}$), the qualitative existence of such an optimum is a general feature. It will naturally arise in any setup where conjugate quadratures are coupled to the same macroscopic observable via nonlinear strong-field dynamics.

%===================================================================
\section{Discussion}
\label{sec:discussion}
%===================================================================

The present results establish a clear operational separation between
classical stochastic effects and covariance-driven quantum signatures
in strong-field dynamics. The ensemble-averaged HHG cutoff shift
[Eq.~\eqref{eq:s_cutoff}] depends only on the marginal amplitude
distribution of the driving field and therefore admits a fully
classical statistical interpretation. Any classical stochastic field
engineered to reproduce the same amplitude statistics would generate
an equivalent mean cutoff response. In contrast, the cutoff variance
inherits the constrained redistribution of fluctuations between
conjugate quadratures imposed by the Heisenberg relation. Within the
single-mode Gaussian framework considered here, the resulting sub-SQL
behavior [Figs.~\ref{fig:quantum_advantage}(c), (d)] cannot be
reproduced within any classical Gaussian field admitting a
non-negative Glauber--Sudarshan $\mathcal{P}$-representation. The
relevant resource is therefore not the amplitude variance alone, but
the full quadrature covariance structure linking conjugate degrees of
freedom. The full angular dependence
$\Delta H^{2}\propto\sigma_{X}^{2}(\theta)$ provides the decisive test:
reproducing this profile simultaneously for all $\theta$ requires
$\det(\Sigma)=1/4$, i.e., saturation of the Robertson--Schrödinger
uncertainty bound, a condition satisfied only by minimum-uncertainty
quantum states, as established in the SM~\cite{SM}.

The normalized variance ratio derived in Eq.~\eqref{eq:ratio}
provides the operational foundation of the proposed witness. At
leading order in the small parameter $\epsilon$, the ratio
$\Delta H^{2}/\Delta H_{\mathrm{SQL}}^{2}=e^{-2r}$ becomes independent
of the absolute vacuum-field amplitude
$\mathcal{E}_{\mathrm{vac}}\propto V_{\mathrm{eff}}^{-1/2}$.
Consequently, the witness remains intrinsically calibrated against the
dominant uncertainty associated with the effective quantization volume
in nanoscale confinement geometries. This distinguishes the present
observable from absolute variance measurements that scale with
$\mathcal{E}_{\mathrm{vac}}^{2}$ and therefore inherit substantial
uncertainty from the poorly defined quantization geometry. Although
higher-order corrections reintroduce explicit dependence on
$\mathcal{E}_{\mathrm{vac}}$, the leading-order scaling remains
intrinsically calibrated.

This robustness also extends naturally to multimode pulse
configurations. Under identical squeezing applied uniformly across all
occupied pulse modes, the normalized ratio $e^{-2r}$ remains invariant
independent of mode number. Deviations from this ideal scaling arise
when the squeezing bandwidth $\Delta\omega_{\rm sq}$ does not fully
cover the pulse bandwidth $\Delta\omega_{\rm pulse}$, leaving
unsqueezed spectral components that contribute additional vacuum noise,
as quantified in the SM~\cite{SM}.

The subleading variance minimum observed near
$r_{\mathrm{opt}}\!\approx\!1.6$ originates from competition between
two independent fluctuation channels coupled to the same macroscopic
observable. Amplitude squeezing suppresses the direct field-amplitude
noise contribution proportional to $e^{-2r}$, while simultaneously
enhancing the conjugate phase-noise channel proportional to
$e^{+2r}$. The resulting turnover is therefore a direct manifestation
of conjugate-quadrature redistribution under the Heisenberg
constraint. Although the precise location of the optimum depends on
the saddle-point structure underlying the ionization-time integral,
its existence is expected to be generic in nonlinear strong-field
observables where both quadratures contribute to the same response.

Several theoretical limitations should also be emphasized. The present
analysis assumes a single-mode Gaussian driver and undepleted
propagation, with coherent phase locking between the squeezed source
and the carrier field. Within this single-mode framework, the
stochastic quadrature pair $(X_i, P_i)$ induces fluctuations in the
carrier-envelope amplitude and phase of an otherwise deterministic
pulse envelope; intra-pulse spectral fluctuations are excluded by
construction. Dynamical electron--field back-action and
propagation-induced mode mixing may perturb the Gaussian covariance
structure underlying the witness mechanism. Likewise, extreme spatial
confinement can suppress macroscopic phase matching and coherent
buildup, potentially requiring extended nanostructured emitter arrays,
e.g., dielectric metasurfaces, for practical signal generation~\cite{vampa2017,liu2018high}. Spatially varying near fields may
additionally introduce dephasing through ponderomotive gradients and
trajectory-dependent phase accumulation~\cite{ciappina2017,husakou2011above}.

Experimentally, the required operating regime necessitates intense squeezed coherent light---a phase-locked superposition of a squeezed vacuum and a strong coherent carrier. Although generating these macroscopic displaced states differs from the zero-mean bright squeezed vacuum (BSV) utilized in several recent strong-field experiments, advances in ultrafast quantum light control~\cite{Sennaryetal2025} suggest this parameter space is becoming accessible at mid-infrared wavelengths. Crucially, because the proposed witness relies exclusively on shot-to-shot spectral statistics, its measurement bypasses the need for complex homodyne detection or direct quadrature reconstruction. Instead, the primary experimental challenge lies in stabilizing the mean driving field; slow drifts in $E_0$ would introduce technical variance that could obscure the predicted sub-SQL scaling. To mitigate this, the simultaneous acquisition of coherent-state reference spectra under identical conditions offers a practical route for common-mode noise rejection and rigorous SQL calibration.

More generally, the mechanism identified here is not restricted to
HHG. A broad class of nonlinear processes that selectively filter field
fluctuations and map quadrature covariance into measurable macroscopic
responses may support analogous witness structures, suggesting a wider
family of covariance-sensitive strong-field observables capable of
operationally probing quantum noise redistribution in nonlinear
optical systems.
%===================================================================
\section{Conclusion}
\label{sec:conclusion}
%===================================================================

In conclusion, we have shown that the shot-to-shot variance of the
HHG cutoff energy, when normalized to its coherent-state value,
descends below the SQL under amplitude squeezed light
driving. This normalized ratio constitutes a Heisenberg witness---an
observable whose sub-SQL scaling certifies the redistribution
of quantum uncertainty between conjugate quadratures---and remains
operationally accessible using standard shot-to-shot spectral
measurements without homodyne detection or photon counting.

Our central result is the leading-order relation
\begin{equation*}
  \frac{\Delta H^{2}(r)}
       {\Delta H_{\mathrm{SQL}}^{2}}
  =
  e^{-2r},
\end{equation*}
which is exact at $\mathcal{O}(\epsilon^{0})$ and independent of the
absolute vacuum-field amplitude and effective mode volume. This
self-calibrating normalization isolates the squeezing-induced
covariance signature from the dominant geometric uncertainties
inherent to nanoscale strong-field geometries, distinguishing the
present observable from absolute variance measurements that scale with $\mathcal{E}_{\mathrm{vac}}^{2}$.

We further identified a subleading phase-noise channel in which the
anti-squeezed conjugate quadrature couples back into the same
macroscopic observable, producing a variance minimum near
$r_{\mathrm{opt}}\!\approx\!1.6$. The existence of this optimum is
generic to nonlinear strong-field processes in which conjugate
quadratures couple into the same observable through distinct
fluctuation pathways, although its precise location remains
model-dependent.

More broadly, these results establish a route for transducing
quantum-optical covariance into experimentally accessible strong-field
observables. The mechanism is not specific to HHG. Any sufficiently
nonlinear process that filters a marginal field distribution while
coupling a conjugate-quadrature channel into the same macroscopic
response---such as nonsequential double ionization or strong-field
photoelectron interferometry---may support an analogous witness
structure. The findings therefore suggest a wider class of
covariance-sensitive observables in nonlinear and ultrafast optics,
emerging wherever the Heisenberg constraint couples 
conjugate noise channels into a common nonlinear response.
%===================================================================
\section*{Data Availability}
%===================================================================
The data that support the findings of this study are available from
the corresponding author upon reasonable request.

%===================================================================
\section*{Author Contributions}
%===================================================================
T.K.\ conceived the theoretical framework, developed and implemented
the stochastic TDSE simulation code, performed all numerical
calculations, and wrote the manuscript. R.T.S.\ and I.L.\ supervised
the project, provided critical guidance on the theoretical
interpretation and manuscript preparation, and reviewed and revised
the manuscript. All authors discussed the results and approved the
final version.

\begin{acknowledgments}
This work was supported by the Australian Research Council through the Discovery Project program (Grant No.~DP230101253).
\end{acknowledgments}

\bibliography{references}
\end{document}